\documentclass[english,,twocolumn]{revtex4}
\usepackage[T1]{fontenc}
\usepackage[latin9]{inputenc}
\usepackage{amsmath}
\usepackage{amssymb}
\usepackage{graphicx}
\usepackage{esint}

\makeatletter

\providecommand{\tabularnewline}{\\}

\@ifundefined{textcolor}{}
{%
 \definecolor{BLACK}{gray}{0}
 \definecolor{WHITE}{gray}{1}
 \definecolor{RED}{rgb}{1,0,0}
 \definecolor{GREEN}{rgb}{0,1,0}
 \definecolor{BLUE}{rgb}{0,0,1}
 \definecolor{CYAN}{cmyk}{1,0,0,0}
 \definecolor{MAGENTA}{cmyk}{0,1,0,0}
 \definecolor{YELLOW}{cmyk}{0,0,1,0}
}

\makeatother

\usepackage{babel}
\begin{document}

\preprint{This line only printed with preprint option}

\title{Rational Chebyshev Spectral Transform for the dynamics of high-power
laser diodes}

\author{J.~Javaloyes}

\email{julien.javaloyes@uib.es}

\affiliation{Universitat de les Illes Balears, C/ Valldemossa, km 7.5, E-07122
Palma de Mallorca, Spain}

\author{S. Balle}

\affiliation{Institut Mediterrani d'Estudis Avançats, CSIC-UIB, E-07071 Palma
de Mallorca, Spain}
\begin{abstract}
This manuscript details the use of the rational Chebyshev transform
for describing the transverse dynamics of high-power laser diodes,
either broad area lasers, index guided lasers or monolithic master
oscillator power amplifier devices. This spectral method can be used
in combination with the delay algebraic equation approach developed
in \cite{JB-OE-12}, which allows to substantially reduce the computation
time. The theory is presented in such a way that it encompasses the
case of the Fourier spectral transform presented in \cite{PJB-JSTQE-13}
as a particular case. It is also extended to the consideration of
index guiding with an arbitrary profile. Because their domain of definition
is infinite, the convergence properties of the Chebyshev Rational
functions allow handling the boundary conditions with higher accuracy
than with the previously studied Fourier method. As practical examples,
we solve the beam propagation problem with and without index guiding:
we obtain excellent results and an improvement of the integration
time between one and two orders of magnitude as compared with a fully
distributed two dimensional model. 
\end{abstract}
\maketitle

\section{Introduction}

An increasing demand for high-power and high-brightness laser diodes
\cite{CEW-JSTQE-13} stems from applications such as solid state and
fiber laser pumping, telecommunications, remote sensing, medicine
or material processing. Laser diodes offer high wall-plug efficiency,
reliability, long lifetime, relatively low investment costs and a
small footprint, but their output power is limited because of Catastrophic
Optical Damage (COD) of the facets. The most direct path to increase
the output power is to reduce the power density on the facets, which
can be achieved by increasing the lateral size of the diode up to
several hundreds of $\mu$m. These so-called Broad-Area Laser Diodes
(BALDs) have allowed to obtain output powers $\sim10$W in CW operation
\cite{Crump_2012}. The increase in power, however, usually implies
a degradation of the optical quality of the beam: the emission profile
in the lateral dimension varies with current due to thermal lensing
and spatial hole burning in the carrier density. These phenomena result
in high $M^{2}$ factors that limit the ability to focus the beam
\cite{Leidner_2012}, and they might even lead to chaotic filamentation
\cite{Thompson_1972,Lang_1993} of the beam. In fact, when several
high order modes are usually present, the emission profile is usually
not stationary \cite{Fischer_1996}.

In order to achieve high-output power with an improved beam quality
from laser diodes, two main approaches have been pursued, the monolithic
Master-Oscillator Power Amplifier (MOPA) and the Tapered Laser (TL),
although alternative options do exist \cite{Schultz_2010,Lang_1998,Paschke_2003,Adachihara_1993,MartinRegalado_1996,Mandre_2005}.
Such TLs have an active region of varying width along the propagation
direction \cite{Odriozola_2009}; usually, their design includes a
relatively short ($\sim500\,\mu$m long) straight region that supports
a single lateral mode with a longer ($\sim1-2$ mm) section whose
width expands significantly, reaching $\sim100\,\mu$m at the output
facet. The straight region is intended to provide spatial mode filtering,
but the longitudinal mode structure of the device is defined by the
total length of the optical cavity. The structure of MOPA \cite{Spreemann_2009}
devices is similar to TLs. They integrate a low-power, conventional
single-mode laser diode (which provides an optical beam of good quality)
with a semiconductor optical amplifier for boosting the power level;
in this case, however, the longitudinal mode spectrum is intended
to be determined by the short section only, which requires using anti-reflection
coatings of very high quality on the output facet. In both cases,
one of the major challenges is the design of the connection of the
two sections in order to minimize the onset of multiple peaks in the
lateral far and near field profiles, specially at high currents, where
the optical spectrum also tends to broaden. Often, angled index steps
known as beam spoilers are included in this region in order to improve
the spatial filtering effect of the short straight region.

The complexity of these devices has stimulated the development of
sophisticated models and simulation tools that can guide the design
of these devices for improving their performances \cite{HK2-PRA-96,Hess_I_2008,LSL-JSTQE-09}.
These models must include not only the structure of the passive cavity
(through the spatial distribution of dielectric constants) but also
the state of the active region through the material's response function
that depends on the spatial distribution of carriers and temperature
\cite{Bandelow_Thermo}. Indeed, it is crucial to reach a self-consistent
solution for the electrical, thermal and optical problems in order
to correctly describe the behavior of these devices; however, the
different time scales involved in these processes may allow to treat
each of them separately, imposing self-consistency at the end. 

The optical simulation module has to solve for the field and carrier
distribution in the device. The three-dimensional nature of the problem
is often reduced to only the lateral and longitudinal directions by
invoking the effective index approximation in the dimension perpendicular
to the active region plane. Time-independent models based on Helmholtz
equation have been developed in order to determine the lateral mode
structure of these devices. In this approach, no a priori assumptions
are made about either the field profile or its characteristics, and
the problem is solved on a spatial grid with sub-wavelength resolution;
given the large size of the device, this is a very memory-expensive
approach. A less-demanding approximation is provided by Beam-Propagation
Methods \cite{Lang_1991,Siegman_2000a,Blaaberg_2007}, where the longitudinal
field profile is decomposed into unidirectional traveling waves, the
amplitudes of which are assumed to vary slowly along the cavity. This
Slowly-Varying Amplitude Approximation (SVA) allows to relax the discretization
requirements, and provides a good approximation to the modal profile
under single-mode operation, including self-focusing of the beam.
However, time-independent models cannot deal with dynamical phenomena,
like beam filamentation, nor predict the onset of instabilities. For
this reason, several time-dependent models have been developed. Most
of these models are also based on the decomposition of the intra-cavity
field into counter-propagating waves \cite{F-PRL-68,F-PRB-70,JB-JQE-09,JB-JQE-10,Freetwm}
in the SVA, differing mainly in the description of the optical response
of the material, hence they are generically termed as Traveling-Wave
Models (TWMs) \cite{JB-JQE-09,JB-JQE-10,Freetwm}.

The numerical approaches for tackling these problems can be classified
as Finite-Difference (FD) methods and spectral or pseudo-spectral
methods. FD methods easily allow for the use of non-uniform grids,
transparent or radiative boundary conditions can be readily implemented,
and they are accurate for high index-contrast geometries. However,
the Von Neumann analysis reveals that only the central part of the
spectrum, i.e. the shallowest frequencies, are properly advected during
the dynamics while another (large) part of the spectrum suffers from
severe amplitude and phase deformations, see \cite{NR-BOOK} Chap.
20. Instead, spectral methods do not distort the spectrum because
they exactly consider the spatial derivative operators; as a consequence,
when seeking for high accuracy smooth solutions, like e.g. the transverse
modal structure of a BALD, spectral and pseudo-spectral algorithms
may converge to the solution much faster than FD algorithms, due to
the so-called \textquotedblleft{}infinite order\textquotedblright{}
or \textquotedblleft{}exponential\textquotedblright{} convergence
\cite{Boyd89a} of the former.

The most popular spectral method is certainly the Fourier Transform
(FT) whose main advantages are to diagonalize the derivative operators
and to be achievable via an algorithm of low complexity ($N\log N$,
where $N$ is the number of points in the spatial grid \cite{CT-MC-65})
and to be readily available as a quality open source software \cite{FFTW3}.
In addition, the method of Exponential Differencing \cite{C-JCP-02}
combined with the FT spectral method allows to treat exactly the influence
of the spatial derivatives operators like diffusion and diffraction
over an interval that does not need to be infinitesimal (i.e. say,
relatively large temporal or spatial steps can be used). It has been
recently shown \cite{PJB-JSTQE-13} that such an exact treatment of
diffraction over a large increment can work in synergy with the delay
Algebraic Equation (DAE) mapping of TWMs developed in \cite{JB-OE-12}.
Such a DAE mapping typically allows for a reduction of the number
of degrees of freedom ---hence the computation time--- between one
and two orders of magnitudes, which alleviates the need of complex
parallel codes for the simulation of BALD, MOPA and TL devices.

However, the FT spectral method automatically imposes periodic boundary
conditions in the lateral direction that are wrong for laser diodes,
where the physical boundary conditions for the field amplitude are
of the radiative type even for the laterally confined modes that decay
exponentially away from the active region. Hence, one is in general
forced to consider a sufficiently large transverse domain in which
the field in only concentrated in the central stripe and as such,
possesses enough space to decay ``naturally'' to zero and not feel
the presence of the wrong boundary conditions. Tapered devices are
particularly demanding from this point of view: the need to extend
the current strip only over a small part of the numerical domain has
to be satisfied even at the exit facet, where the tapered device has
its maximal breadth, which then defines the lateral size of the rectangular
box. However, an acceptable resolution where the tapered device is
the broadest means that at the same time the thinner part may only
be resolved by a handful of points. As such, one is forced to use
an extremely demanding discretization everywhere in order to properly
resolve the narrower part and one ends up with an extremely large
rectangular numerical domain composed of a very fine grid in which
the active current stripe extend only over a fraction of the rectangle.
The inclusion of an absorbing Perfectly Matched Layer (PML) close
to the boundary may may help to mitigate this problem; however, an
abrupt transverse variation of the losses would in principle give
rise to a convolution product in the Fourier domain which may render
the inclusion of such a PML if not difficult, at least costly from
a numerical point of view. Another drawback of the FT spectral method
is that its basis functions (plane waves) are not the most adequate
for describing the exponential decaying tails of the field outside
of the central stripe, which may result in a sub-optimal convergence.

The aforementioned points suggest that a spectral method based upon
a set of functions which decay, even weakly, in $x=\pm\infty$ may
improve on the convergence properties, as clearly indicated in the
discussion at the end of \cite{W-NM-92}. In addition, a non uniform
discretization ---with a high density of points in the central stripe
where the non linear dynamics occurs, and a low density of points
in the outer regions, where the field decays linearly--- would represent
a clear improvement. These considerations hint toward the use of a
different spectral method, namely the Rational Chebyshev Transform
(RCT). The RCT devised by Christov consists in a modification of the
Chebyshev Transform defined on $x\in[-1,1]$ onto $x\in[-\infty,\infty]$
by an arctangent mapping \cite{C-JAM-82}. The basis consists in a
full set of rational fraction function which tends to zero at $x=\pm\infty$
and that are represented over a non uniform mesh of points whose density
decreases with the distance from the origin. Such a method solves
the aforementioned drawback of the FT method while keeping essentially
its good properties, i.e. a simple representation of the spatial derivative,
even upon Exponential Differencing \cite{C-JCP-02} , and surprisingly
enough, a fast $N\log N$ implementation based upon the Fourier transform.

In this work, we discuss how a RCT spectral method over a variable
grid following the variations of the current stripe profile along
the cavity axis can be implemented for a TWM at a marginal increase
in complexity as compared to the FT method developed in our earlier
work \cite{PJB-JSTQE-13}. We use the TWM developed in previous works\cite{Freetwm}
generalized to the presence of transverse diffraction and including
both index- and gain-guided sections as encountered in the different
sections of MOPAs and TLs. Propagation along the axis is solved in
time-domain using the DAE formalism \cite{JB-OE-12}, which enables
using a coarse discretization along the optical axis. The numerical
algorithm is formally equivalent for both the RCT and the FT approaches.
As such, we present our theory in an unified way where only the precise
form of the matrices corresponding to the second derivative, the index
guiding and the extra term due to the variable grid differ between
the RCT and FT method.

This manuscript is organized as follows. In Section II, we recall
the basis of our TWM \cite{Freetwm} generalized to the presence of
transverse diffraction. We detail in Section III the basic properties
of the RCT, and in Section IV the representation in this basis of
the second order spatial derivatives and of transverse index guiding.
Section V is devoted to the implementation of the DAE transformation
using Exponential Differencing \cite{C-JCP-02} and discuss how one
must treat the boundary conditions along the longitudinal direction.
We detail in particular the many caveats present in the evaluation
of the integration weights due to the stiffness incurred by the spectral
transformation. We also show how the full two dimensional mesh profile
can be reconstructed from the few actives ``slices'' that remain
active with the DAE approach. In Section VI, we exemplify the validity
of our approach by studying a variety of cases and comparing the FT
and the RCT methods.

\section{Model}

Our model considers a single mode structure in the transverse direction,
which is reduced via the effective index approximation. The optical
field in the cavity is assumed to be almost TE polarized and it is
decomposed into a forward and a backward wave of SVA $E_{\pm}(x,z,t)$.
In addition, the carrier density $N(x,z,t)$ in the cavity is decomposed
into a quasi-homogeneous term $N_{0}(x,z,t)$ and a (weak) grating
term at half the optical wavelength, $N_{\pm2}(x,z,t)$, with $N_{-2}(x,z,t)=N_{2}^{*}(x,z,t).$
The instantaneous distributions $E_{\pm},$ $N_{0}$ and $N_{\pm2}$
in the lateral ($x\in[-\infty;\infty]$) and longitudinal ($z\in[0;L_{z}]$)
direction are described in the paraxial approximation by a TWM \cite{Freetwm}
extended to include diffraction in the transverse dimension. From
the numerical point of view, however, the lateral size of the region
is taken as large but finite, i.e., $x\in[-L_{x}/2;L_{x}/2]$. Scaling
the two spatial coordinates and time as $(x,z,t)=(X,Z,T)/(L_{x},L_{z},\tau)$
, where the time of flight in the cavity is $\tau=nL_{z}/c$, the
resulting equations read 
\begin{eqnarray}
\left(\partial_{t}\pm\partial_{z}\right)E_{\pm} & = & i\Delta_{0}\partial_{x}^{2}E_{\pm}+\left[i\psi\left(x,z\right)-\alpha_{i}\right]E_{\pm}+iP_{\pm}\;,\label{eq:1}\\
\partial_{t}N_{0} & = & J\left(x,z\right)-R\left(N_{0}\right)+\mathcal{D}\partial_{x}^{2}N_{0}\label{eq:2}\\
 & - & i\left(P_{+}E_{+}^{\star}+P_{-}E_{-}^{\star}-c.c.\right),\nonumber \\
\partial_{t}N_{\pm2} & = & -\left[R'\left(N_{0}\right)+4\mathcal{D}_{0}q_{0}^{2}\right]N_{\pm2}\label{eq:3}\\
 & - & i\left(P_{\pm}E_{\mp}^{\star}-E_{\pm}P_{\mp}^{\star}\right),\nonumber \\
R\left(N\right) & = & AN+BN^{2}+CN^{3}.\label{eq:4}
\end{eqnarray}

\noindent where $\alpha_{i}$ takes into account for the internal
losses that we assume constant for the sake of simplicity, $\psi\left(x,z\right)\in\mathbb{R}$
describes the lateral index guiding profile in the weak guiding approximation.
The spatially dependent pump current profile is $J(x,z)$. For the
sake of simplicity, we restrict our analysis to case where the current
stripe possesses a constant transverse width, index and current profiles,
i.e. $J(x,z)=J\left(x\right)$ and $\psi(x,z)=\psi\left(x\right)$.
The optical wave vector along the propagation direction $z$ in the
medium is $q_{0}=2\pi n/\lambda_{0}$, and we have defined the scaled
diffusion and diffraction lengths are $\mathcal{D}=\tilde{\mathcal{D}}L_{x}^{-2}$
and $\Delta_{0}=\lambda_{0}/(4\pi n_{g})L_{z}L_{x}^{-2}$ respectively,
with $\lambda_{0}$ the wavelength in vacuum and $n$ and $n$$_{g}$
being the effective index and the effective group index. The scaled
index guiding profile $\psi$ is related to the waveguide structure
by the relation $\psi\left(x,z\right)=\left(q_{0}L_{z}\right)\left(\delta n\left(x,z\right)/n_{g}\right)$
and it is typically an order one quantity even when $\delta n\sim10^{-3}$.
For the sake of simplicity we have neglected the transverse diffusion
( i.e. $\mathcal{D}\partial_{x}^{2}N_{2}$) of the longitudinal population
grating $N_{2}$ at half the optical wavelength $\lambda_{0}$ since
the actual decay rate $4\mathcal{D}q_{0}^{2}\sim10^{12}\,$s$^{-1}$
corresponds to a length well below the diffusion length. The non linear
recombination $R\left(N\right)$ is assumed of a cubic form with the
fitting parameters $A,B$ and $C$, and contains contributions from
several sources as the non radiative, the radiative and the Auger
contributions. The boundary conditions at the left and right facets
read in the simplest case of a Fabry-Perot cavity 
\begin{equation}
\negthickspace E_{+}\left(x,0,t\right)\negthickspace=\negthickspace r_{l}E_{-}\left(x,0,t\right),E_{-}\left(x,1,t\right)\negthickspace=\negthickspace r_{r}E_{+}\left(x,1,t\right).
\end{equation}
The current stripe is defined between $x_{\pm}=\pm\frac{1}{2}$. The
discretization in the longitudinal direction is simple and as detailed
in \cite{PJB-PRA-10} we define an uniform mesh along the direction
$z$ composed of $N_{z}$ points separated by a distance $h=N_{z}^{-1}$,
located at abscissae $z_{j}$ shifted of $h/2$ from the boundaries
$z=\left(0,1\right)$ and defined by 
\begin{eqnarray}
z_{j} & = & h\left(j-\frac{1}{2}\right)\quad j\in[1,N_{z}].
\end{eqnarray}

With this approach, the boundary conditions are applied at half integer
time steps as discussed in the appendix of in \cite{PJB-PRA-10}.

When the dynamics is restricted to the vicinity of the band-gap, as
it is the case of BALDs, it is possible to use a Padé approximation
to the optical response as in \cite{JB-JQE-09,PJB-JSTQE-13}. Upon
transforming back to time domain using that $\partial_{t}\rightarrow-i\omega$,
the time domain evolution equation for the polarizations can be written
at a given point ($x,z$) as 
\begin{eqnarray}
\partial_{t}P_{\pm} & = & -\left(\frac{1}{ib}+i\omega_{p}\right)P_{\pm}+\frac{\chi-\omega_{p}a}{ib}E_{\pm}+\frac{a}{b}\partial_{t}E_{\pm}\nonumber \\
 & + & \frac{\partial_{N}\chi}{ib}N_{\pm2}E_{\mp}.\label{eq:11}
\end{eqnarray}

\noindent where $\chi=\chi(N_{0}(x,z,t),\omega_{p})$ and $a$ and
$b$ are given in \cite{JB-JQE-09,PJB-JSTQE-13} respectively, with
$N=N_{0}(x,z,t)$. The spatial dependence of $a$ and $b$ arises
from that of $N\left(x,z\right)$. If not otherwise stated we use
the parameters described in Table.1. 
\begin{table}
\begin{tabular}{|c|c|c||c|}
\hline 
Symbol  & Value  & Units  & Meaning\tabularnewline
\hline 
$\lambda_{0}$  & $1.55$  & $\mu$m  & Emission wavelength \tabularnewline
\hline 
$n$  & $3.75$ & -  & Effective index \tabularnewline
\hline 
$\Gamma$  & $5\%$  & -  & Optical confinement factor \tabularnewline
\hline 
$R$  & $1/0.5/0.01$  & -  & Facet power reflectivity\tabularnewline
\hline 
$\tau_{c}$  & $12.5$  & ps  & Cavity transit time\tabularnewline
\hline 
$2\alpha_{i}$  & $15$  & cm$^{-1}$  & Internal losses\tabularnewline
\hline 
$\Omega_{g}$  & $0$  & GHz & Band-Edge Frequency \tabularnewline
\hline 
$2\chi_{0}$  & $1500$  & cm$^{-1}$  & Gain factor\tabularnewline
\hline 
$\gamma$  & $8\times10^{12}$  & s$^{-1}$ & Polarization decay rate\tabularnewline
\hline 
$N_{t}$  & $1\times10^{18}$  & cm$^{-3}$  & Carrier Density at transparency\tabularnewline
\hline 
$\mathcal{D}$  & $20$  & cm$^{2}$s$^{-1}$  & Ambipolar diffusion coefficient\tabularnewline
\hline 
$A$  & $1\times10^{8}$  & s$^{-1}$ & Non radiative recombination\tabularnewline
\hline 
$B$  & $7\times10^{-10}$  & cm$^{3}$s$^{-1}$  & Spontaneous recombination\tabularnewline
\hline 
$C$  & $1\times10^{-29}$  & cm$^{6}$s$^{-1}$  & Auger recombination\tabularnewline
\hline 
\end{tabular}

\caption{Table of the parameters used in the simulations}

\label{Table1} 
\end{table}

\section{Rational Chebyshev Transform\label{sec:Rational-Chebyshev-Transform}}

Assuming that the field and the polarization decay to zero in $x=\pm\infty$,
and following the notations of \cite{W-NM-92}, we define the finite
Rational Chebyshev Transform of order $N_{x}$ (with $N_{x}=2N$ an
even number with the notations of \cite{W-NM-92}) of the field and
of the polarization in the transverse direction as 
\begin{eqnarray}
E_{\pm}\left(x,z,t\right) & = & \sum_{n=-N_{x}/2}^{N_{x}/2-1}\tilde{E}_{\pm}\left(n,z,t\right)\rho_{n}\left(x\right)\,,\\
P_{\pm}\left(x,z,t\right) & = & \sum_{n=-N_{x}/2}^{N_{x}/2-1}\tilde{P}_{\pm}\left(n,z,t\right)\rho_{n}\left(x\right)\,,
\end{eqnarray}
where the basis set is defined by the complex rational fractions 
\begin{eqnarray}
\rho_{n}\left(x\right) & = & \frac{\left(ix-1\right)^{n}}{\left(ix+1\right)^{n+1}}.
\end{eqnarray}

The link between the RCT and the Fourier transform can be revealed
by using change of variables $x=\cot\left(\theta/2\right)$ which
maps the entire real axis $x\in[-\infty,\infty]$ onto $\theta\in[0,2\pi]$.
Noticing that 
\begin{eqnarray}
\rho_{n}\left(x\right) & = & \exp\left(in\theta\right)/\left(1+ix\right)\,,
\end{eqnarray}
we can emphasis the direct link with the Fourier Transform since
\begin{eqnarray}
\left(1+ix\right)E_{\pm}\left(x,z,t\right) & = & \sum_{n=-N_{x}/2}^{N_{x}/2-1}\tilde{E}_{\pm}\left(n,z,t\right)e^{in\theta}.\label{eq:RCT}
\end{eqnarray}

Consequently, the RCT transform of a function $E\left(x\right)$ defined
over $N$$_{x}$ grid points, the Right Hand Side (RHS) of Eq.~\ref{eq:RCT},
can be obtained by simply seeking the discrete Fourier transform of
the function $\left(1+ix\right)E\left(x\right)$, that is to say the
left hand side (LHS) of Eq.~\ref{eq:RCT}, provided that the function
$E\left(x\right)$ is evaluated over the specific non uniform grid
as defined by the $x_{j}=\cot\left(\theta_{j}/2\right)$, with an
uniform distribution of the phases $\theta_{j}$ as 
\begin{eqnarray}
\theta_{j} & = & 2\pi j/N_{x}\quad,\quad j\in[0,N_{x}-1].
\end{eqnarray}
This justifies why the RCT is often called a ``Fourier transform
in disguise'' \cite{Boyd89a}. A great wealth of information on the
Chebyshev Transform and the RCT on infinite and semi-infinite intervals
can be found in the free resource \cite{Boyd89a}. Notice that the
infinity is included as a point $x_{0}=\infty$ and therefore the
limiting behavior of the function must be imposed during the transformation,
i.e. $\left(1+ix_{0}\right)E\left(x_{0}\right)=\lim_{x\rightarrow\infty}\left(1+ix\right)E\left(x\right)=0.$
As detailed in \cite{C-JAM-82}, the completeness and orthogonality
of the basis set $\rho_{n}\left(x\right)$ follows from the properties
of the Fourier Transform. In the rest of this manuscript we will be
representing the spectral transformation of $E$ (let it be either
the RCT or the FT) as a tilde operator, i.e. $\tilde{E}$.

\begin{figure}
\begin{centering}
\includegraphics[bb=0bp 0bp 382bp 283bp,clip,width=9cm]{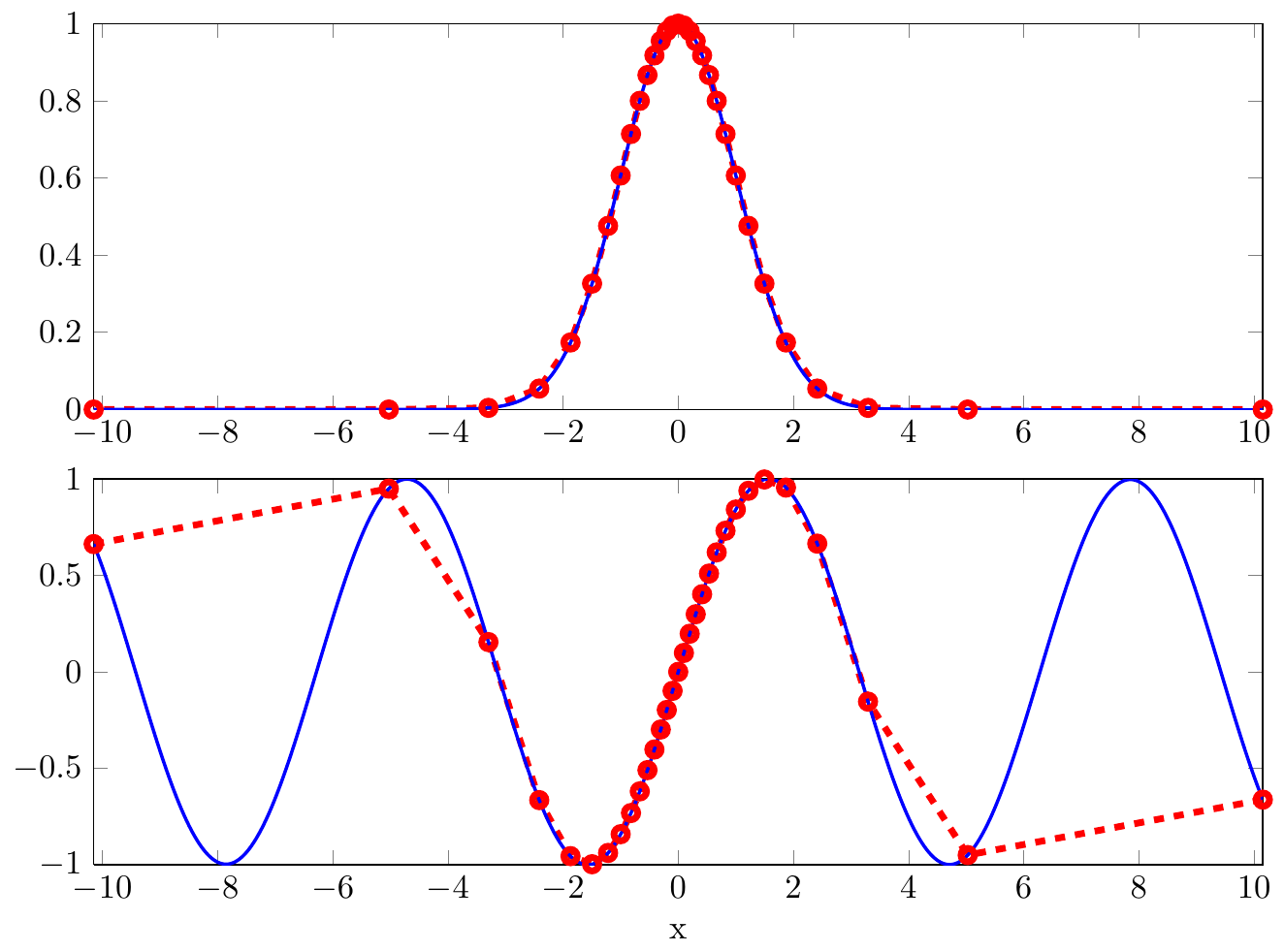}
\par\end{centering}

\caption{Representation of a localized function $\exp\left(-x^{2}/2\right)$
and a delocalized wave $\sin\left(x\right)$ over the non uniform
RCT grid width $N_{x}=32$.}

\label{fig_ex_rct}
\end{figure}

Here the advantage provided by RCT for the problem at hand is that
it represents more accurately than the FT, for a given order $N_{x}$,
the functions that are localized around $x=0$ and which decay smoothly
in $x=\pm\infty$. This is due to the fact that the density of mesh
points decays with the distance from the origin, see Fig.~\ref{fig_ex_rct}.
Since we know beforehand that we are going to deal exclusively with
such decaying and localized functions, due to the fact that the electrical
pumping is localized only over the central stripe, the RCT will prove
superior to the FT which is more adapted to represent ``delocalized''
functions over the whole domain. But it is only by trading the possibility
to represent accurately such delocalized functions that we obtain
a superior resolution in the vicinity of the current stripe. A neat
example of the huge improvement provided in these cases by the RCT
is found in Fig.~1 of \cite{TJ-ANZ-08}, which compares the convergence
of the RCT and of the FT methods in the case of the 1+1 Non Linear
Schrödinger equation. Here, the numerical solution obtained with the
RCT is indistinguishable from the exact hyperbolic secant with only
a handful of mesh points ($N_{x}=32$) , while the FT method fails
qualitatively, by predicting an off-centered, symmetry breaking solution. 

\section{index guiding and diffraction}

In order to simplify the notations we will omit the longitudinal and
temporal dependence whenever possible. Hence the spectral transformation
in the transverse direction of a variable $A\left(x,z,t\right)$ will
be denoted $\tilde{A}$ instead of $\tilde{A}$$\left(n\vert q,z,t\right)$
with $n$ the order of the eigenfunction $\rho_{n}$ for the RCT or
$q$ the spatial frequency of the Fourier mode. This has the added
benefit of allowing to treat both spectral transformations simultaneously.
For the RCT, the fundamental relation for the product of the $\rho_{n}$
disclosed by \cite{C-JAM-82} reads 
\begin{eqnarray}
\rho_{n}\left(x\right)\rho_{k}\left(x\right) & = & \frac{\rho_{n+k}\left(x\right)-\rho_{n+k+1}\left(x\right)}{2}.
\end{eqnarray}
a relation that is very similar to the product of two planes waves
in the case of the FT. Hence we can deduce a relation for the RCT
of the product of two functions akin to a cyclic convolution between
the RCT of the two original functions, 
\begin{equation}
\widetilde{\psi E}=\mathbf{P}\tilde{E}\quad,\quad\mathbf{P}=\frac{\text{\ensuremath{\mathbf{C_{\psi}}}}-\text{\ensuremath{\mathbf{C'_{\psi}}}}}{2}.\label{eq:convolRCT}
\end{equation}

The operation can be constructed as a matrix-vector multiplication,
where the RCT of the function $\psi\left(x\right)$ is converted into
$\left[\text{\ensuremath{\mathbf{C_{\psi}}}}-\text{\ensuremath{\mathbf{C'_{\psi}}}}\right]/2$.
The matrix $\text{\ensuremath{\mathbf{C_{\psi}}}}$ is a circulant
Toeplitz matrix whose lines are composed of the Fourier transform
of the function $\left(1+ix\right)\psi\left(x\right)$ and $\mathbf{C_{\psi}'}$
corresponds to a circular shift of all the columns of $\text{\ensuremath{\mathbf{C_{\psi}}}}$
to the left, or identically a circular shift down of all the lines.
In the case of a FT, we would simply have $\mathbf{P}=\text{\ensuremath{\mathbf{C_{\psi}}}}$
but with the lines of $\text{\ensuremath{\mathbf{C_{\psi}}}}$ composed
by the Fourier transform of the function $\psi\left(x\right)$. Assuming
that the index profile $\psi\left(x\right)$ besides being a real
valued function is symmetric with respect to the center to the current
stripe entails that $\mathbf{P}=\mathbf{^{t}P}$ which allow for some
optimization and to replace half of the multiplications involved in
evaluating Eq.~\ref{eq:convolRCT} by additions. 

With respect to other representations, the FT has the obvious advantage
that in the basis of the plane waves, the second derivative operator
is diagonal, which justifies the success and the efficiency of the
so-called split-step methods. In \cite{PJB-JSTQE-13}, we exploited
fully this fact since it allowed us to integrate exactly the traveling
wave equations along the spatio-temporal characteristics $s_{\pm}=z\mp t$.
There, we could describe exactly the effect of the differential operator
$\exp\left[i\left(\Delta_{0}h\right)\frac{\partial^{2}}{\partial x^{2}}\right]$
for relatively large values of the step ($h\sim0.1$) using the so-called
Exponential Differentiation. This allowed the decimation method \cite{JB-OE-12}
to be extended to the case of transverse diffractive dynamics. In
the RCT basis, the second derivative operator is not that simple and
following the derivation of \cite{W-NM-92} (notice a typo in the
original derivation of \cite{C-JAM-82}) we have that 
\begin{eqnarray}
\frac{\partial^{2}}{\partial x^{2}}E_{\pm}\left(x,z,t\right) & = & \sum_{n=-N_{x}/2}^{N_{x}/2-1}\tilde{E}_{\pm}^{\left(2\right)}\left(n,z,t\right)\rho_{n}\left(x\right)\,,
\end{eqnarray}
 with the following expression of the coefficients $\tilde{E}_{\pm}^{\left(2\right)}\left(n\right)$ 

\begin{align}
\tilde{E}_{\pm}^{\left(2\right)}\left(n\right) & =-\frac{1}{4}\left[n\left(n-1\right)\tilde{E}_{\pm}\left(n-2\right)-4n^{2}\tilde{E}_{\pm}\left(n-1\right)\right.\nonumber \\
 & +\left(6n^{2}+6n+2\right)\tilde{E}_{\pm}\left(n\right)-4\left(n+1\right)^{2}\tilde{E}_{\pm}\left(n+1\right)\nonumber \\
 & \left.+\left(n+1\right)\left(n+2\right)\tilde{E}_{\pm}\left(n+2\right)\right],
\end{align}
that is to say 
\begin{align}
\tilde{E}_{\pm}^{\left(2\right)}\left(n,z,t\right) & =\mathbf{M}\tilde{E}_{\pm}\left(n,z,t\right),
\end{align}
where $\mathbf{M}$ is the second order differentiation matrix in
the Chebyshev space. Comparing this result to the more standard spectral
method based on the FT, we have here a symmetric penta-diagonal matrix
$\mathbf{M}$ instead of a diagonal one. Taking the exponential of
the second order derivative operator as in \cite{PJB-JSTQE-13} over
a segment of length $h$ defines the exponential differential operator
$\mathbf{w_{M}}=\exp\left[i\left(\Delta_{0}h\right)\mathbf{M}\right]$
which is composed of two block diagonal matrices of size $N_{x}/2$.
At this point, one would think that the possible improvements one
obtained by using the RCT instead of the FT are going to be mitigated
by the fact that taking the second order derivative has a much increased
computational cost. Indeed, such cost would be here of $N_{x}^{2}/2$
multiplications, instead of $N_{x}$ in the case of the FT method.
However, a plot of the values of the matrix elements of $\mathbf{w_{M}}$
is instructive. 

\begin{figure}
\begin{centering}
\includegraphics[bb=0bp 0bp 318bp 171bp,clip,width=9cm]{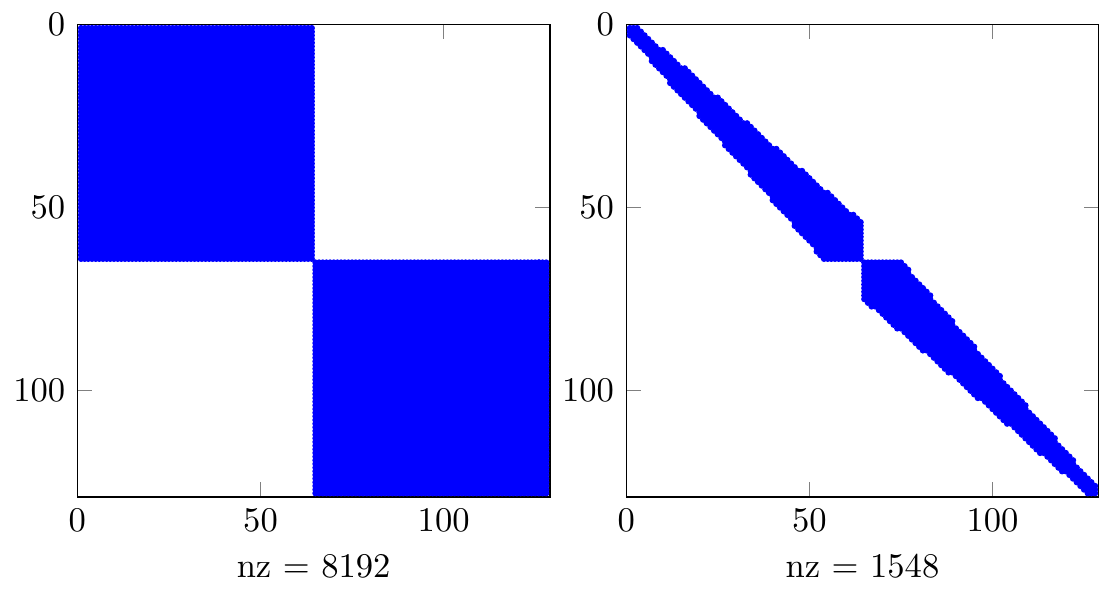}
\par\end{centering}

\caption{Representation of the exponential of the second order derivative $\mathbf{w_{M}}=\exp\left[i\left(\Delta_{0}h\right)\mathbf{M}\right]$
with $\Delta_{0}h=3.3\times10^{-4}$ (equivalent to a $200\,\mu$m
wide current stripe and a large longitudinal step of $100\,\mu$m)
(left) and the same matrix (right) suppressing elements of square
modulus smaller than $10^{-8}$. The number of non zero elements $nz$
is strongly reduced. Here $N_{x}=128$.}

\label{fig_ex_M2sparse}
\end{figure}
On notice in Fig.~\ref{fig_ex_M2sparse} that the elements of $\text{\ensuremath{\mathbf{w_{M}}}}$
decrease quickly with respect to the diagonal. By enforcing $w_{M}^{ij}=0$
whenever $w_{M}^{ij}<\varepsilon$, we recover a matrix whose structure
is banded and symmetric. Such a nice property of the finite second
derivative operator $\mathbf{w_{M}}$ stems from a physical reason.
In a BALD $\left(h\Delta_{0}\right)\ll1$ and an expansion of $\mathbf{w_{M}}$
in powers of $\mathbf{M}$ converges quickly, which explains the limited
bandwidth. In other worlds, diffraction is only a weak perturbation
of the wave propagation in a broad current stripe, which explains
that even upon quite large a propagation distance (i.e. exponential
differencing) the diffraction operator conserves its bandwidth limited
form. Notice that the sparsity of $\mathbf{w_{M}}$ was not exploited
in \cite{TJ-ANZ-08} where the exponential of $\mathbf{M}$ was explicitly
considered as a full matrix.

\begin{figure}
\begin{centering}
\includegraphics[bb=0bp 0bp 426bp 226bp,clip,width=9cm]{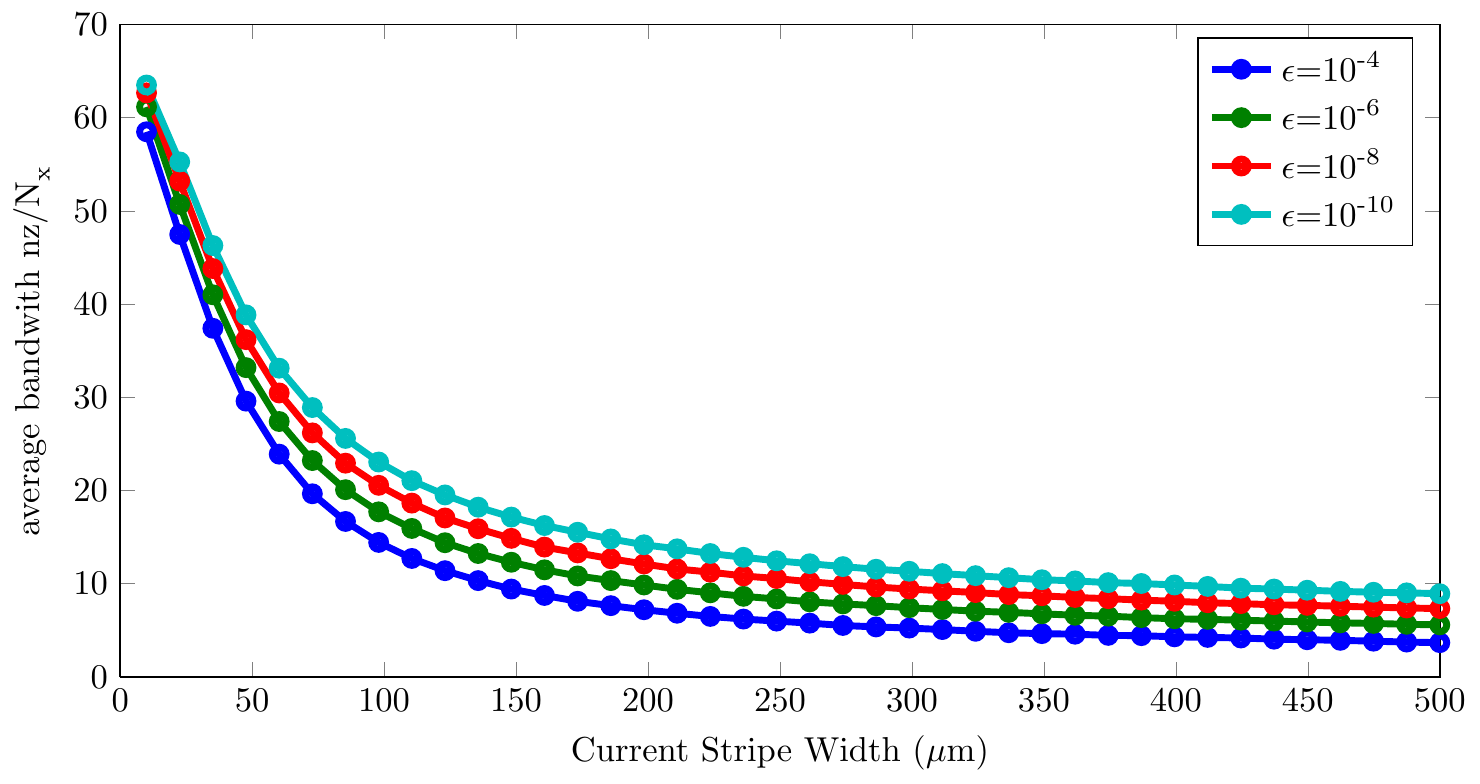}
\par\end{centering}

\caption{Representation of the average bandwidth of $\mathbf{w_{0}}$ as defined
by the number of elements whose modulus square is larger than $\varepsilon$
divided by the number of lines, for various values of $\varepsilon$.
Here $N_{x}=128$.}

\label{fig_ex_bandwidth}
\end{figure}

We represent in Fig.~\ref{fig_ex_bandwidth} the evolution of the
average bandwidth of $\mathbf{w_{M}}$ as a function of the magnitude
of the effect of yjr diffraction $\left(\Delta_{0}h\right)$. Typically,
for a large longitudinal step, i.e. a length of the characteristic
of $100\,\mu$m ($h=0.1$ with the parameters of Section V) and an
uniform current stripe of width $2L_{x}$ of $125,\,250$ and $500\,\mu$m,
i.e. $h\Delta_{0}=8.4\times10^{-4},\,2.1\times10^{-4}$ and $5.2\times10^{-5}$,
the average bandwidth of $\mathbf{w_{M}}$ is $17,\,10$ and $7$
with a cut-off value $\varepsilon=10^{-8}$. Notice that the total
multiplication cost involved shall further be divided almost by two
using the symmetry $\mathbf{w_{M}=\mathbf{^{t}w_{M}}}$, which makes
in practice the RCT method almost equivalent in speed as the FT for
the operation of taking the exponential of the second derivative.
The RCT would get less and less efficient when the current stripe
gets narrower.Here, one would think that the FT becomes more adapted.
However in such a narrow stripe situation, one would certainly consider
a strong index guiding thereby destroying anyway the banded structure.
This would also happen with the FT method. Hence, the RCT remains
a good choice in all practical situations. 

\begin{figure}
\begin{centering}
\includegraphics[bb=0bp 0bp 406bp 222bp,clip,width=9cm]{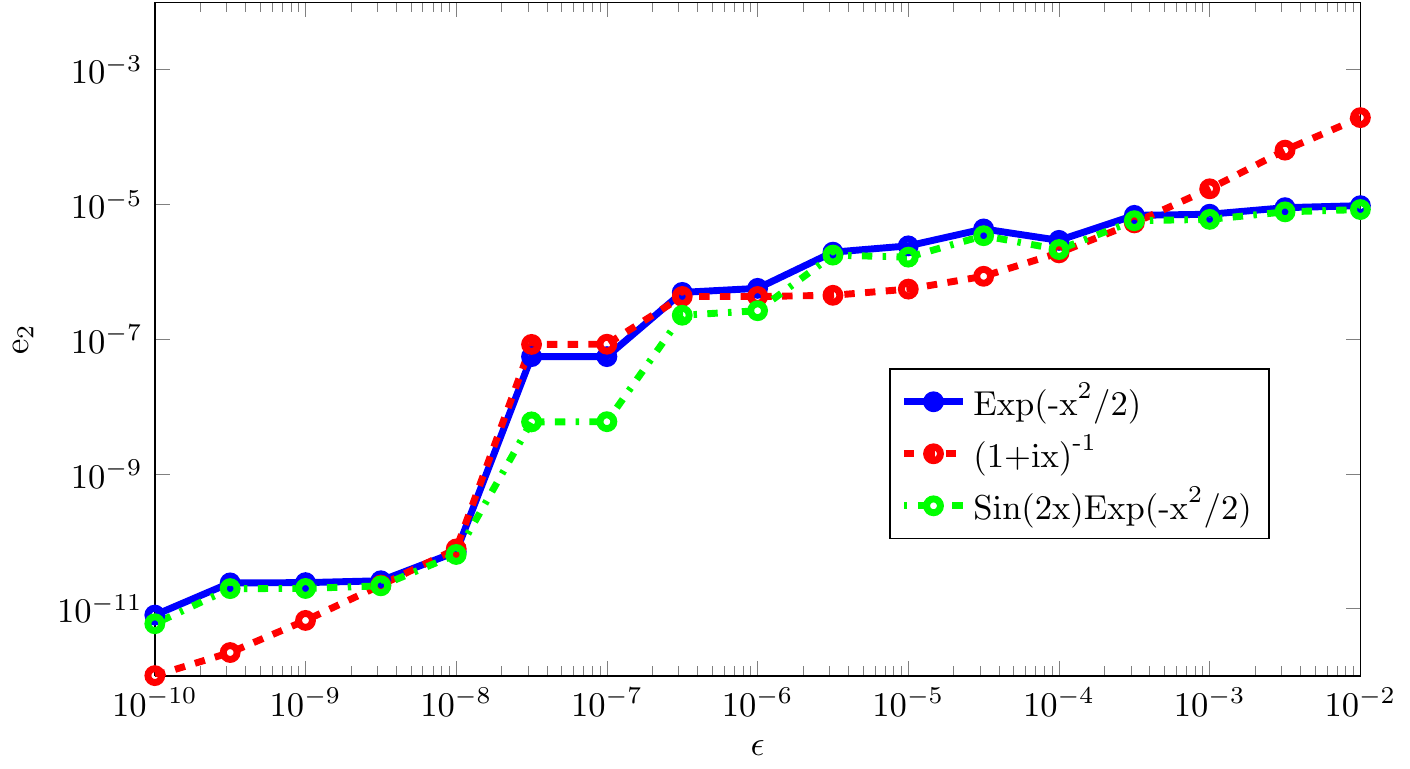}
\par\end{centering}

\caption{Integral of the square modulus of the difference between the exact
and approximated exponential of the second derivative for various
functions as a function of the cut-off parameter $\varepsilon$. Here
$N_{x}=128$ and $h\Delta_{0}=3.3\times10^{-4}$.}

\label{fig_d2_error}
\end{figure}

Still, the incurred error due to matrix truncation must remain under
control. We discuss in Fig.~\ref{fig_d2_error} the error in the
operation of taking the exponential of the second derivative due to
the $\varepsilon-$truncation of $\mathbf{w_{M}}$. We represented
$e_{2}$ as defined by 
\begin{eqnarray}
e_{2} & = & \int_{-\infty}^{\infty}dx\left|\mathbf{w_{M}}f\left(x\right)-\mathbf{w_{M}^{\varepsilon}}f\left(x\right)\right|^{2}
\end{eqnarray}

One notice for instance in Fig.~\ref{fig_ex_bandwidth} that the
value of $10^{-8}$ retained previously for $\varepsilon$ induces
an error of $10^{-10}$. As expected, such error is of similar order
for various functions having similar lateral Full Width at Half Maximum
(FWHM) extension.

\section{Delay Algebraic Equation Mapping}

\noindent Our method for solving Eqs.~(\ref{eq:1}-\ref{eq:4}) adapts
the DAE approach developed in \cite{JB-OE-12} for the case of a gain
guided BALD where we treated exactly the free field equation. Such
a method where one integrates exactly some linear contributions has
been used in many contexts, sometimes with a great deal of intuition
in laser physics as in \cite{F-PRL-68}. It is formally termed Exponential
Time Differencing (ETD) \cite{C-JCP-02}, although in our case integration
is carried out in space-time. The ETD allows to obtain an excellent
accuracy through the exact treatment of the stiff linear integrating
factors. Here, although the original problem in its PDE form is not
stiff \emph{per se} along the characteristics, some difficulties are
introduced by the spectral transformation. Mainly, the spectral transformation
introduces stiffness. This is easier to spot in the simplest case
of FT, where the Nyquist mode has an spatial evolution scale proportional
to the inverse of its frequency which grows with the transverse mesh
size, i.e. $q_{N}^{2}=\left(\pi/\delta x\right)^{2}\sim N_{x}^{2}$.
The RCT also gives rise to a second order differentiation matrix of
similar stiffness since the eigenvalues of $\mathbf{M}$ increase
quadratically from a very small to some very large number, see for
instance the discussion regarding the first Figure of \cite{W-NM-92}.
It is why we shall treat exactly the contribution of the stiff linear
terms. Such analytical treatment consists in integrating the free
field equation over the so-called forward and backward characteristics
\cite{F-PRB-70} while performing a linear approximation of the variation
of the source term $\tilde{P}_{\pm}$ which is at the same time a
small and a smooth perturbation of the free field propagation, another
formulation of the so-called Uniform Field Limit approximation of
Laser physics \cite{NA-BOOK-78}. 

In a normal spatio-temporal TWM, one must use an identical temporal
and longitudinal increment, i.e. $h=\delta t$ in order to fulfill
the CFL condition \cite{CFL}. The DAE approach allows to decouple
these two values and to ``leapfrog'' between spatial points, hence
the term of mesh decimation. Such a decoupling is relevant since the
temporal increment must be chosen according to the \emph{temporal}
stiffness of the active material response, while the spatial increment
is related to the gain amplification and to the field \emph{non uniformity}
along the characteristics. By analogy with a TWM approach, we define
the decimation factor $D$ denoting the number of skipped spatial
points as $D=h/\delta t$. When $D=1$, one recovers the usual TWM
, see \cite{JB-OE-12} for details.

In the absence of source $P_{\pm}=0$, the general form of the spectral
form of the wave equations in the presence of losses and index guiding
reads
\begin{eqnarray}
\left(\partial_{t}\pm\partial_{z}\right)\tilde{E}_{\pm}\left(n\vert q,z,t\right) & = & \mathbf{Q}\tilde{E}_{\pm}\left(n\vert q,z,t\right)
\end{eqnarray}
with the following definition of $\mathbf{Q}$
\begin{equation}
\mathbf{Q}=i\Delta_{0}\mathbf{M}+i\mathbf{P}-\alpha_{i}\mathbf{Id}.
\end{equation}
and of $\mathbf{w_{0}}\left(z\right)$ the free field propagator over
an interval $z$
\begin{eqnarray}
\mathbf{w_{0}}\left(z\right) & = & \exp\left(z\mathbf{Q}\right)\label{eq:w0-1}
\end{eqnarray}

Here, we recall that the same formalism can be used either for the
RCT or the FT, simply replacing $\mathbf{M}$ by the diagonal matrix
of the second derivative in the FT basis ${\rm \left(\frac{\pi}{2}\right)^{2}diag}\left([-N_{x}/2,...,N_{x}/2+1]\right)^{2}$
and the index guiding matrix $\mathbf{P}$ by $\mathbf{C}$.

\noindent In the presence of an active source, we start by factoring
out the influence of the free field propagator by defining 
\begin{eqnarray}
\tilde{E}_{\pm} & = & \mathbf{w_{0}}\left(\pm z\right)U_{\pm}
\end{eqnarray}
hence the wave equation becomes
\begin{equation}
\left(\partial_{t}\pm\partial_{z}\right)U_{\pm}=\mathbf{w_{0}^{-1}}\left(\pm z\right)i\tilde{P}_{\pm}=\mathbf{w_{0}}\left(\mp z\right)i\tilde{P}_{\pm}.
\end{equation}
whose integral solution over a characteristics of length $h$ is
\begin{eqnarray}
\tilde{E}_{\pm}\left(n,z,t\right) & = & \mathbf{w_{0}}\left(h\right)\tilde{E}_{\pm}\left(n,z\mp h,t-h\right)+i\mathcal{S}_{\pm},\label{eq:DAE1}
\end{eqnarray}
with the following definition of the source $\mathcal{S}_{\pm}$ 
\begin{eqnarray}
\mathcal{S}_{\pm} & = & \int_{0}^{h}\mathbf{w_{0}}\left(h-s\right)\tilde{P}_{\pm}\left(n\vert q,z\mp h\pm s,t-s\right)ds.\label{eq:DAE2}
\end{eqnarray}

As in \cite{PJB-JSTQE-13}, the source term $\mathcal{S}_{\pm}$ is
approximated by assuming a linear variation of the polarization along
the characteristics, i.e. we assume that the distance $h$ between
the two points is short enough for the UFL to hold. As such we have,
up to third order in $h$ 
\begin{align}
\mathcal{S}_{\pm} & \simeq\mathbf{w_{1}}P_{\pm}\left(n,z\mp h,t-h\right)\nonumber \\
 & +\mathbf{w_{2}}\left[P_{\pm}\left(n,z,t\right)-P_{\pm}\left(n,z\mp h,t-h\right)\right]\label{eq:pola2}
\end{align}
with 
\begin{eqnarray}
\mathbf{w_{1}} & = & \int_{0}^{h}ds\, e^{\left(h-s\right)\mathbf{Q}},\label{eq:w1}\\
\mathbf{w_{2}} & = & \int_{0}^{h}ds\, e^{\left(h-s\right)\mathbf{Q}}\left(\frac{s}{h}\right).\label{eq:w2}
\end{eqnarray}
The temporal evolution of the field and polarization is therefore
given by
\begin{eqnarray*}
\tilde{E}_{\left(\pm\right),j}^{l+1} & = & \mathbf{w_{0}}\tilde{E}_{\left(\pm\right),j\mp1}^{l}+i\mathbf{w_{1}}\tilde{P}_{\left(\pm\right),j\mp1}^{l}\\
 & + & i\mathbf{w_{2}}\left(\tilde{P}_{\left(\pm\right),j}^{l+1}-\tilde{P}_{\left(\pm\right)j\mp1}^{l}\right),
\end{eqnarray*}

For the evolution of the forward (resp. backward) field on the first
(resp. last) point, located at a distance $h/2$ from the boundaries
$z=0$ and $z=1$ respectively, the coefficients must be integrated
over half an interval which is obtained simply by replacing $h$ by
$h/2$ in the equations (\ref{eq:w0-1},\ref{eq:w1},\ref{eq:w2}).
The practical evaluation of the expressions given by Eqs.~(\ref{eq:w1},\ref{eq:w2})
is discussed in the appendix.

\subsection{Boundary conditions }

\begin{figure}
\begin{centering}
\includegraphics[bb=0bp 30bp 392bp 630bp,clip,width=9cm]{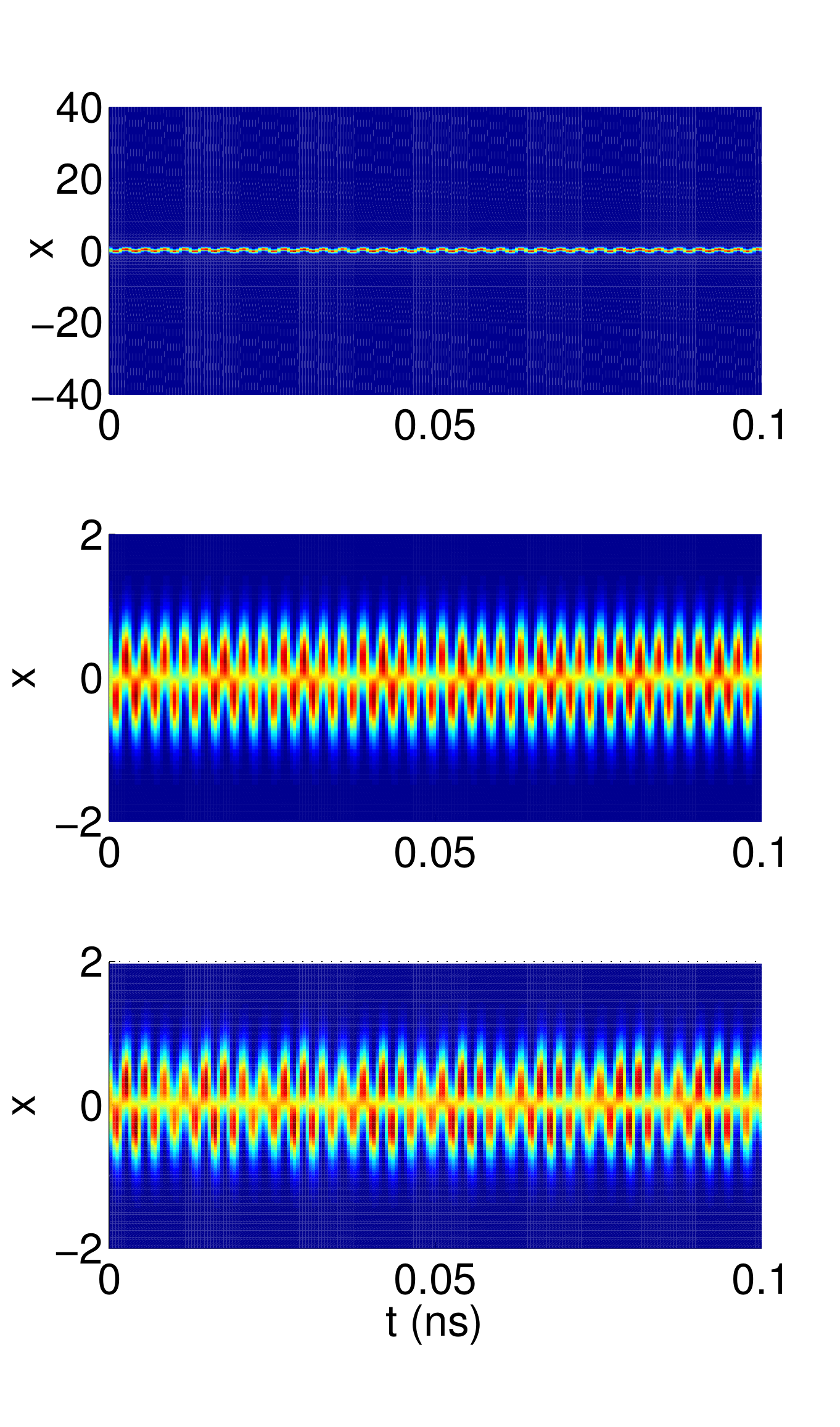}
\par\end{centering}

\caption{Temporal time trace of the right facet output in the case of a $28\,\mu$m
wide BALD. A zipper state emerge and remains stable up to four times
threshold, the current value here is $2.5\, J_{th}$. The time traces
correspond to the near field intensity pattern on the right facet,
i.e. $\left|E_{+}\left(x,1,t\right)\right|^{2}$ after a $25\,$ns
simulation time. The top line corresponds to the full domain with
the RCT method, while the central panel is a zoom on the central region
of the stripe. An excellent agreement between the RCT approach (top/center)
and the case of the FT method (bottom) is found but in this latter
case, the full numerical domain is shown. In all cases, the parameters
are $N_{x}=128$ and $D=16$. The mirror reflectivity drops to $0$
outside the region $x\in[-1,1]$. }

\label{fig_diaphragm}
\end{figure}

Since the field are propagated toward the boundaries in direct space,
it is trivial to introduce reflection and or transmission coefficient
that depend on the longitudinal coordinate in order for instance to
model the effect of a diaphragm on the cavity. We present the result
of such an approach in Fig.\ref{fig_diaphragm} where the reflectivity
drops to $0$ outside of the stripe defined in the interval $x\in[-1;1]$.
The dynamics obtained here is very similar to the zipper state present
in the case of gain guided devices. Due to the drop of reflectivity
outside of the central region, the FT method performs here as well
as the RCT since the high losses induced by the mirror reflectivity
profile helps in this particular case confining the field in the central
region. 

\subsection{Spatio-temporal numerical algorithm}

One remarks that the DAE can only be written in the spectral space
which is not convenient for the active material description. Hence
the methodology consists in transforming back and forth between the
direct and the Chebyshev/Fourier space to update the field and the
polarization and carriers with the following sequence akin to the
pseudo-spectral methods 
\begin{enumerate}
\item Perform a first order explicit Euler prediction of $\tilde{E}_{\pm}\left(n\vert q,z,t+\delta t\right)$
and transform back to space ($x$) domain to obtain a first order
estimate of $E_{\pm}\left(x,z,t+\delta t\right)$.
\item Perform a second order semi-implicit evolution of the polarization
and get $P_{\pm}\left(x,z,t+\delta t\right)$ as in \cite{JB-JQE-09}
and perform the spectral transformation in order to get $\tilde{P}_{\pm}\left(n\vert q,z,t+\delta t\right)$.
\item Perform the second order semi-implicit evaluation of $\tilde{E}_{\pm}\left(n\vert q,z,t+\delta t\right)$
as in \cite{JB-JQE-09} and transform back to space ($x$) domain
to get $E_{\pm}\left(x,z,t+\delta t\right)$.
\item By using a staggered grid for the population inversion (see the appendix
of \cite{PJB-PRA-10} for instance), the carrier equations eqs.~(\ref{eq:2}-\ref{eq:3})
are treated as in \cite{JB-JQE-09}. Here, the effect of carrier diffusion
although not critical must be treated in the spectral space due to
the non uniform grid if one uses the RCT.
\end{enumerate}

\section{Example and Convergence properties}

The longitudinal length of all devices is fixed to the value of $1\,$mm
which leads to a single trip of $\tau_{c}=12.5\,$ps and to a longitudinal
mode spacing of $40\,$GHz. The temporal increment is $\delta t=\tau_{c}/N_{z}\sim48\,$fs.
In order to test the numerical method described in the previous section,
we consider three BALDs of width $W_{x}=28\,\mu$m, $W_{x}=57\,\mu$m
and $W_{x}=114\,\mu$m. As noted before the lateral extent of the
integration domain in the case of the FT is $L_{x}=2W_{x}$ while
for the RCT it is much larger and increases with $N_{x}$ as $\sim2N_{x}/\pi$
while keeping $50\%$ of the mesh point inside a domain of width $W_{x}$.
We assume typical values for the semiconductor material as detailed
in Table 1. In addition, we fixed the longitudinal mesh size to be
$N_{z}=257$. Such a convenient value allows us to use decimation
factors $D=(128,64,32,16,8,4,2,1)$. In addition, if one use power
of two for $N_{z}-1$ and $D$, for increasing values of $D$ each
mesh is a subset of the previous one which allows restarting the simulation
from the previous ones to study the accuracy of the truncation for
increasing values of $D$. 

\begin{figure}
\begin{centering}
\includegraphics[bb=30bp 25bp 600bp 620bp,clip,width=9cm]{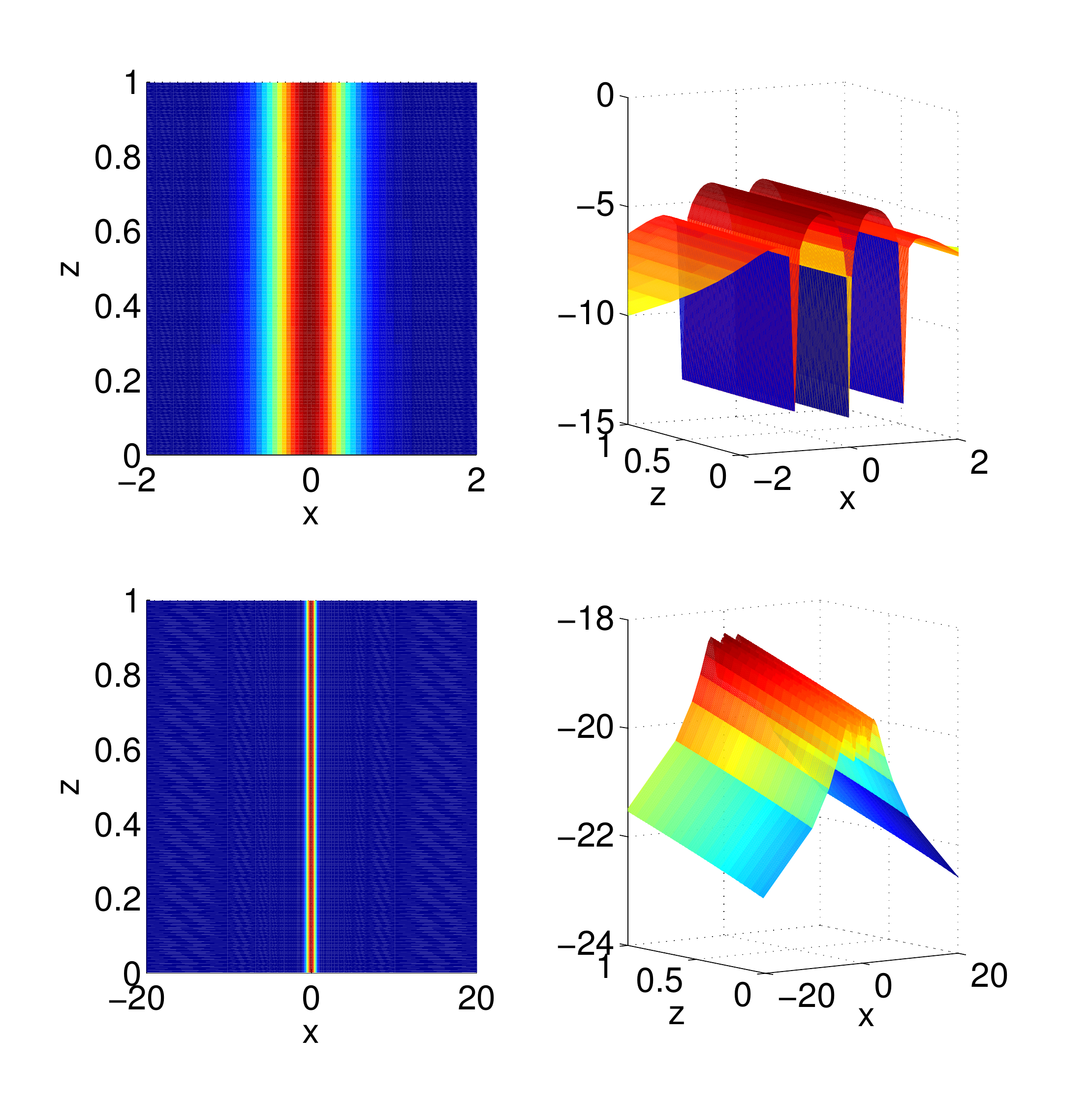}
\par\end{centering}

\caption{Numerical solution (left) after a single round-trip and deviation
(right) from the analytical solution for the Fourier (top) and the
Chebyshev (bottom) transforms. For clarity the $\log_{10}$ of the
error is represented. Close to the domain boundaries the error is
is $10^{-6}$ and $10^{-21}$ for the FT and the RCT, respectively.
The horizontal interval $[-1,1]$ corresponds to $W_{x}=57\,\mu$m.}

\label{fig_error_diffraction-1}
\end{figure}

As an initial condition, we use as a transverse profile a Gaussian
beam defined as
\begin{eqnarray}
G_{a}\left(x,z\right) & = & \frac{1}{\sqrt{z-ia}}\exp\left[-\frac{ix^{2}}{4\Delta_{0}\left(z-ia\right)}\right].\label{eq:Gauss}
\end{eqnarray}
Notice that there is no mistake regarding the presence of the square
root in Eq.(\ref{eq:Gauss}), the standard result in the literature
being the solution to the paraxial equation in two dimensions. The
main advantage of the Gaussian beam is that it is an eigenfunction
of the paraxial equation in the absence of non linear gain and index
guiding. Therefore we can use such a profile to compare the numerical
and analytical results and study the convergence of both the FT and
RCT methods after some integration time. Indeed, if the initial condition
is 
\begin{eqnarray}
E_{+}\left(x,z,0\right) & = & G_{a}\left(x,z\right)\exp\left(ikz\right)
\end{eqnarray}
with $k=m\pi-i\ln\left(r_{l}r_{r}\right)/2$, a wave-vector solution
of the longitudinal propagation problem, after an integer number of
round-trip $n$ and therefore a propagation time/length $2n$, the
final solution is simply 
\begin{eqnarray}
E_{+}\left(x,z,2n\right) & \negthickspace\negthickspace=\negthickspace\negthickspace & G_{a}\left(x,z+2n\right)\left(r_{l}r_{r}\right)^{n}\exp\left(-2n\alpha_{i}\right).
\end{eqnarray}
Here, the advantage of the RCT with respect to the FT method will
become quite obvious. We choose as a parameter for the Gaussian beam
$a=10$, which correspond to a well localized profile whose FWHM is
equal to $3/4$. Such initial condition was integrated over a single
round-trip, i.e. $n=1$. With $N_{x}=64$ the error or the FT method
with respect to the analytical result is of the order of $10^{-9}$.
This is a nice result but looking at how the error is distributed
is useful. In Fig.~(\ref{fig_error_diffraction-1}), one notices
that the field does not reach zero on the lateral side: even if the
initial beam was well localized in the cavity, some light will always
leak on the sides and experience the folding incurred by the periodic
boundary conditions imposed by the FT method. Because of that, increasing
$N_{x}$ does not change the value of the error, preventing any further
improvement in the convergence.

\begin{figure}
\begin{centering}
\includegraphics[bb=0bp 0bp 386bp 198bp,clip,width=9cm]{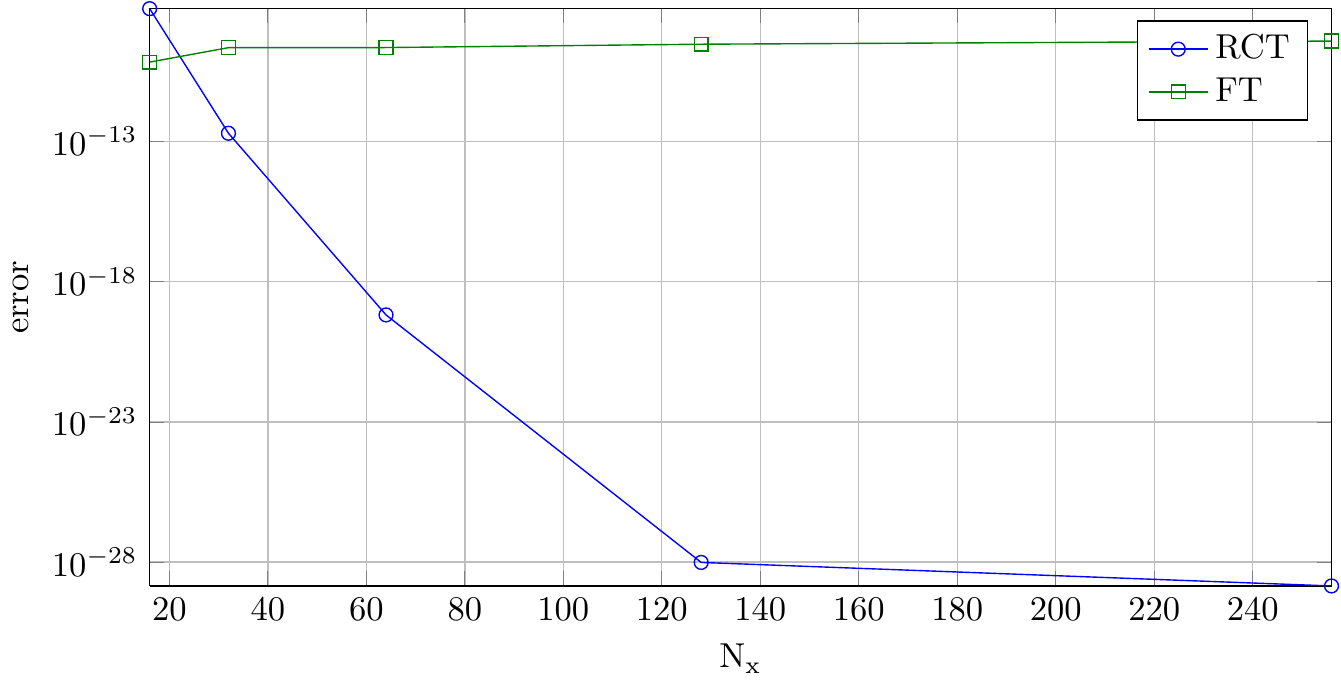}
\par\end{centering}

\caption{Integral of the square modulus of the error between the exact and
the numerical solution after a single trip in the cavity starting
with a Gaussian transverse profile with $a=10$ as an initial condition.
Increasing the mesh size does not improve on the accuracy of the FT
method due to the non vanishing value of the field close to the periodic
boundary condition. On the contrary, the RCT exhibits exponential
convergence due to the linear increase of the numerical domain with
$N_{x}$. }

\label{fig_error_diffraction-2}
\end{figure}

On the other hand, the RCT already gives an error of $10^{-19}$ with
$N_{x}=64$. This is because in this case the numerical domain is
already much broader. Here, the error reduces to $10^{-26}$ with
$N_{x}=128$ signaling exponential convergence. This is due to the
fact that the numerical domain lateral extension increases with the
RCT as $2N_{x}/\pi$. This explain the widely different behavior we
represent in Fig.~(\ref{fig_error_diffraction-2}) regarding the
convergence of the FT and the RCT methods.

\subsection{Reconstruction }

\begin{figure}
\begin{centering}
\includegraphics[bb=30bp 0bp 610bp 314bp,clip,width=9cm]{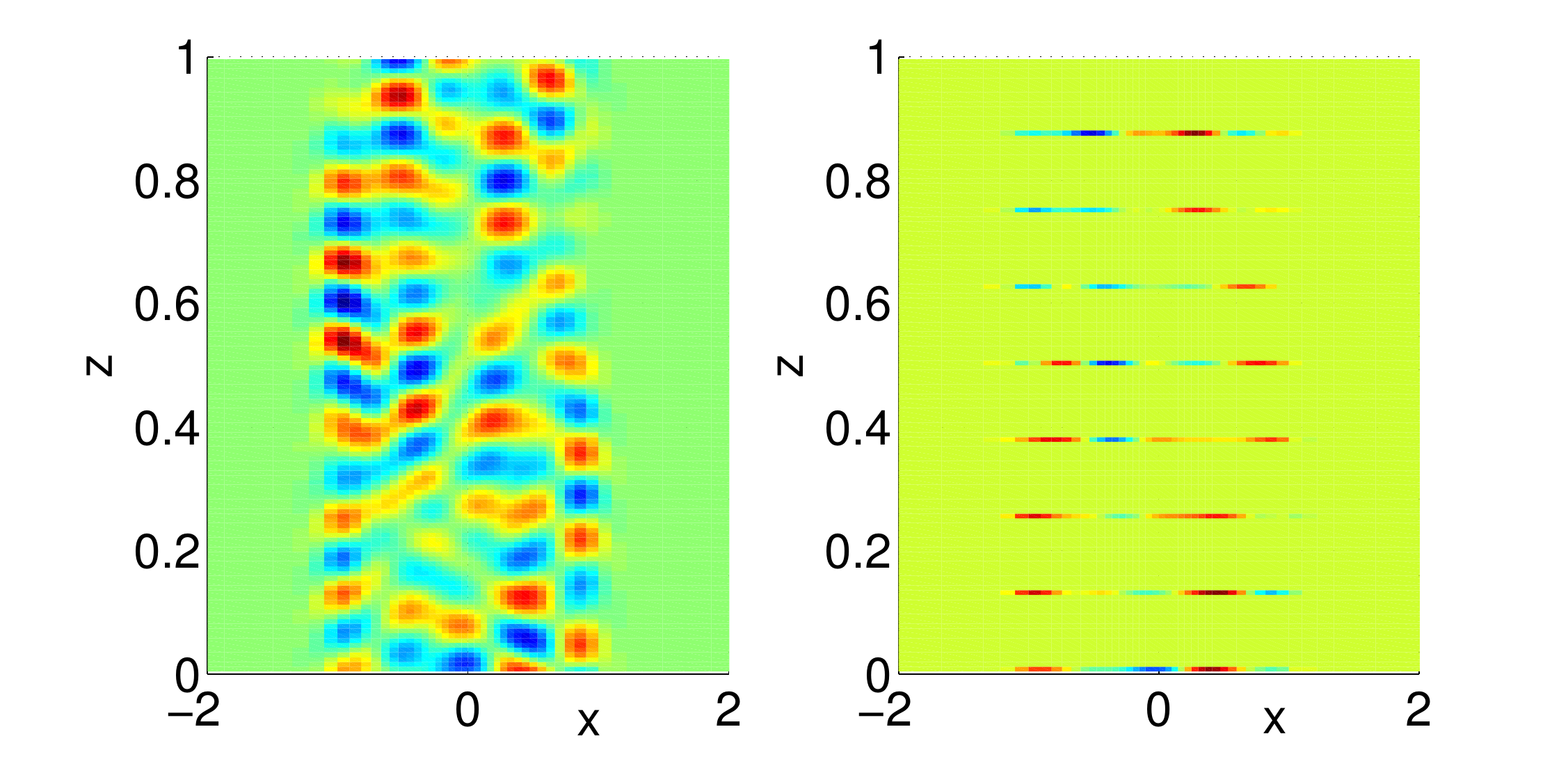}
\par\end{centering}

\caption{Mesh reconstruction from the past values in the case of a $500\,\mu$m
long and $57\,\mu$m wide device in presence of strong index guiding
$n_{0}=3$. Such device support up to $11$ confined modes. The decimation
factor is $D=16$ which corresponds to $(N_{z}-1)/D=8$ effective
active mesh points. The current is $J=5J_{th}$ and the facet reflectivies
are $\left(r_{l},r_{r}\right)=\left(1,1\right)$. Left: real part
of the forward field reconstructed from the only $8$ actives slices
(right) in the case of the RCT. Here $N_{x}=64$ and the horizontal
axis is limited to the central region while the domain extend up to
$x\sim\pm20$. }

\label{fig_reconstruction}
\end{figure}

In the case of a large decimation factor the effective mesh along
the laser cavity propagation axis is only composed of a few transverse
slices. Still, all the complexity of the dynamics remains intact and
is hidden in the past values of the field kept at each transverse
slice. For instance, it is still possible to reconstruct the full
two dimensional profile by using the past values of the fields in
order to say, have a intuitive idea of the two dimensional beam profile.
Such a reconstruction at a point $z_{i}$ is done from a past value
$t_{p}$ at the closest point on the left $z_{l}$ (resp. right $z_{r}$)
for the forward (resp. backward) propagating wave and reads 

\begin{eqnarray}
\tilde{E}_{\pm}\left(n\vert q,z_{i},t\right) & = & \mathbf{w_{0}^{\pm}}\left(z_{l,r},z_{i}\right)\tilde{E}_{\pm}\left(n\vert q,z_{l,r},t_{p}\right)\label{eq:euler}\\
 & + & \mathbf{w_{1}^{\pm}}\left(z_{l,r},z_{i}\right)\tilde{P}_{\pm}\left(n\vert q,z_{l,r},t_{p}\right)\nonumber 
\end{eqnarray}
with $t_{p}=\left|z_{i}-z_{l,r}\right|$. This spatial reconstruction
of the longitudinal profiles of the fields achieved in Eq.~(\ref{eq:euler})
is simply an Euler prediction (from the point of view of the source
term) to recover the corresponding ``missing'' transverse slices.
The result of such a reconstruction is exemplified in Fig.~\ref{fig_reconstruction}
in the case of a BALD operated in a chaotic regime and where the dynamics
is multimode in both the longitudinal and transverse dimensions. The
smoothness of the reconstructed profiles indicates \textit{a posteriori}
that no significant information was lost. Here, in the case of a large
decimation factor the effective mesh along the laser cavity is only
composed of a few slices. Notwithstanding, all the complexity of the
dynamics remains intact and is hidden in the past values of the field
kept at each slice point and a reconstruction can be achieved using
the past values of the fields. Such a reconstruction was discussed
in \cite{PJB-JSTQE-13} in the case of the FT method in the absence
of index guiding. We show in Fig.~\ref{fig_reconstruction} that
the same principle applies also to the presence of index guiding and
for the RCT method. The result of such a reconstruction is exemplified
in Fig.~\ref{fig_reconstruction} in the case of a straight, $500\,\mu$m
wide BALD operated in a chaotic regime and where the dynamics is multimode
in both the longitudinal and transverse dimensions. The smoothness
of the reconstructed profiles indicates \textit{a posteriori} that
no significant information was lost. We started from a a noisy initial
condition and the snapshot was obtained after $1.25\,$ns of simulation
time, i.e. during the strongly multimode turn-on transient.

\subsection{Mode with index guiding }

In the absence of source, we have for a mode $E_{+}$
\begin{eqnarray}
E_{+}\left(x,z,t\right) & = & \Lambda\left(x\right)\exp\left[i\left(kz-\omega t\right)\right],
\end{eqnarray}
with $k$ one of the possibly complex wave-vector associated with
the Fabry-Perot longitudinal modal structure as given by the HelmHoltz
equation
\begin{eqnarray}
i\left(k-\omega\right)\Lambda & = & i\left(\Delta_{0}\frac{d^{2}}{dx^{2}}+n_{0}\Theta\left(x\right)\right)\Lambda,\label{eq:Helmholtz}
\end{eqnarray}
with $\Theta$ the transverse index profile. In the three regions
we assume the solution of the form
\begin{eqnarray}
\Lambda\left(x\right) & = & A_{l}\exp\left(q_{o}x\right),\nonumber \\
\Lambda\left(x\right) & = & A_{+}\exp\left(q_{+}x\right)+A_{-}\exp\left(q_{-}x\right),\label{eq:sol_mode}\\
\Lambda\left(x\right) & = & A_{r}\exp\left(-q_{o}x\right),\nonumber 
\end{eqnarray}
hence with $\beta=k-\omega$, replacing the ansatz in Eq.~(\ref{eq:Helmholtz})
gives
\begin{equation}
\beta=\Delta_{0}q_{0}^{2}\quad,\quad\beta=\Delta_{0}q_{\pm}^{2}+n_{0},
\end{equation}
while the field continuity gives in $\pm1$ defines the dispersion
relation
\begin{eqnarray}
1 & = & \frac{q_{+}+q_{0}}{q_{+}-q_{0}}\frac{q_{-}-q_{0}}{q_{-}+q_{0}}\exp\left[2\left(q_{+}-q_{-}\right)\right].\label{eq:disp_modes}
\end{eqnarray}

\begin{figure}
\begin{centering}
\includegraphics[bb=0bp 0bp 368bp 226bp,clip,width=9cm]{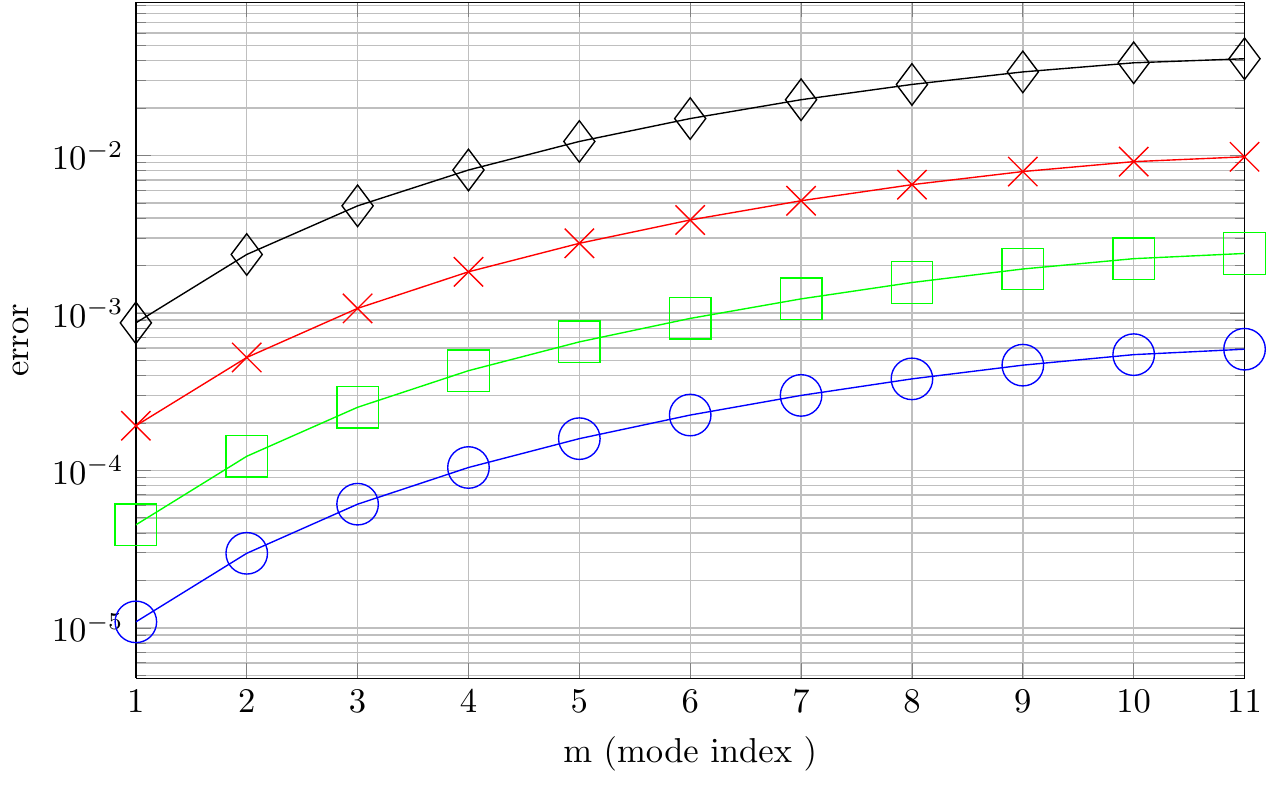}
\par\end{centering}

\caption{Integral of the square modulus of the error between the numerical
and analytical profile for increasing transverse mode index number.
The diamond, cross, square and circles correspond to $N=128,256,512$
and $1024$, respectively. The width of the index guiding region is
$W_{x}=114\,\mu$m.}

\label{fig_error_index_guiding_1}
\end{figure}

Such a non linear eigenvalue problem can also be solved numerically
form a finite size system. Taking the FT or the RCT of Eq.~(\ref{eq:Helmholtz})
yields
\begin{eqnarray}
\left(\Delta_{0}\mathbf{M}+n_{0}\mathbf{P}\right)\tilde{\Lambda} & = & \beta\mathbf{Id}\tilde{\Lambda}.\label{eq:Helm_finite}
\end{eqnarray}
hence one simply needs to diagonalize such a matrix. From the inverse
spectral transform of the eigenvectors, the profiles of the finite
size numerical modes are found. We compare in Fig.~(\ref{fig_error_index_guiding_1})
the difference between the result given by the finite size system
with the analytical solution as defined by Eq.~(\ref{eq:disp_modes}).
In this respect, the eigenvalues of the finite size system as defined
by Eq.~(\ref{eq:Helm_finite}) gives us an excellent guess as a starting
point in order to solve Eq.~(\ref{eq:disp_modes}). The error is
defined as the integral of the square modulus of the difference between
the numerical system of finite size $N_{x}$ and the analytical result
that stem from solving Eq.~(\ref{eq:Helmholtz},\ref{eq:sol_mode},\ref{eq:disp_modes}),
i.e. 
\begin{eqnarray}
e & = & \int_{-\infty}^{\infty}\vert\Lambda_{N_{x}}-\Lambda^{*}\vert^{2}dx
\end{eqnarray}

\begin{figure}
\begin{centering}
\includegraphics[clip,width=9cm]{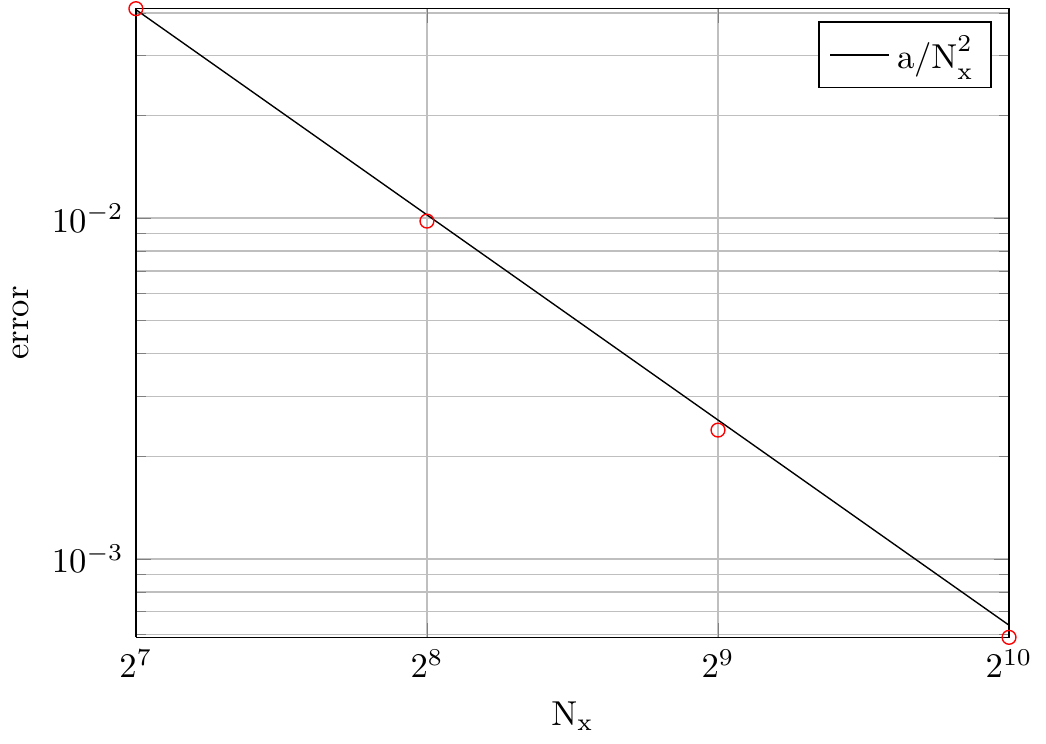}
\par\end{centering}

\caption{Integral of the square modulus of the error between the numerical
and analytical profile for mode number $=11$ as a function of the
number of mesh point $N_{x}$ and best fit with a quadratic law.}

\label{fig_error_index_guiding_2}
\end{figure}

Inspecting the result in Fig.~(\ref{fig_error_index_guiding_1})
indicates that the error increases with the mode number which is reasonable
due to the larger number of oscillations of the transverse profile
for high index values. However, the value of the error with respect
to the analytical solution is much larger than for the case where
only diffraction is present. This is a well known fact and stem from
the presence of a discontinuity. Here, the abrupt variation of the
refractive index profile impedes the spectral method of achieving
exponential convergence. A separate plot of the convergence properties
of a single mode as depicted in Fig.(\ref{fig_error_index_guiding_2})
shows only quadratic convergence which is still a good result.

We used large values of the refractive index jump ($n_{0}=3$) in
order to have a large number of confined modes (here $m_{max}=11$).
As such the beam profiles are well localized in the center of the
numerical domain which explains why we obtained very similar results
regarding the convergence properties and the error values for both
the FT and the RCT methods. It is however foreseeable that the FT
method would be inferior in the case of weak guiding.

\section{Conclusion}

We presented in this manuscript an unified description that allows
using both the Fourier and the Rational Chebyshev spectral method
in combination with the decimation method \cite{JB-OE-12} in presence
of index guiding and non linear gain guiding. We also pointed out
that a non linear adaptive grid can be used with almost no penalty
via the Chebyshev approach. Because they are defined on an infinite
domain the Chebyshev Rational function set allows handling the boundary
conditions with higher accuracy than with the previously studied Fourier
Transform method. This improvement is achieved at a negligible cost
since the Chebyshev transform is merely a Fourier transform in disguise.
However, upon exponentiation of the free field propagator, the Chebyshev
approach has the detrimental effect of leading to full matrices. We
discussed how by inserting a cut-off upon the matrix element values
one can recover a banded sparse matrix. We exemplified our by solving
for the beam propagation dynamics in index guiding and tapered gain
guided configuration. We obtained excellent result and an improvement
of the integration time between one and two orders of magnitudes as
compared with a fully distributed two dimensional method, which may
alleviate in some cases the necessity of using complex parallel codes.
Our approach can be readily extended to other descriptions of the
active medium and for instance to the time domain convolution kernel
recently developed in \cite{JB-PRA-10,JB-JQE-12}. Another important
improvement of the method discussed would consider the inclusion of
the thermal and electro-thermal effects \cite{uwe} due to current
injection and field two photon absorption which are known to play
a dominant role in the dynamics of BALDs. Last but not least, distributed
feedback in the weak coupling approximation and assuming uniformity
of the coupling in the transverse plane can by readily implemented
as in \cite{Freetwm}.

\section*{Acknowlegment }

J.J. acknowledges fruitful discussions with A. Perez-Serrano as well
as financial support from the Ramon y Cajal fellowship. J.J. and S.B.
acknowledge project funding from project RANGER (TEC2012-38864-C03-01)
and from the Direcció General de Recerca, Desenvolupament Tecnològic
i Innovació de la Conselleria d\textquoteright{}Innovació, Interior
i Justícia del Govern de les Illes Balears co-funded by the European
Union FEDER funds.

\subsection*{Appendix}

An analytical expression can be obtained for the integration weights
$\mathbf{w_{1}}$ and $\mathbf{w_{2}}$, 
\begin{eqnarray}
\mathbf{w_{1}} & = & \mathbf{Q}^{-1}\left(\mathbf{w_{0}}-\mathbf{I}\right),\label{eq:w1-1}\\
\mathbf{w_{2}} & = & \mathbf{w_{1}}-\mathbf{Q}^{-1}\left(\mathbf{w_{0}}-\frac{\mathbf{w_{1}}}{h}\right).\label{eq:w2-1}
\end{eqnarray}

Notice that in the worst case with no losses and no index guiding,
the matrix $\mathbf{Q}$ which reduces to $\mathbf{M}$ does not possess
any zero eigenvalue, see theorem 6 for (29) in \cite{W-NM-92}, which
always ensure the existence of full rank inverse $\mathbf{Q}^{-1}$
and the validity of Eqs.~(\ref{eq:w1-1},\ref{eq:w2-1}) in the case
of the RCT. The Eqs.~(\ref{eq:w1-1},\ref{eq:w2-1}) are also well
behaved in the case with the FT but a special treatment must be considered
for the case of the zero eigenvalue corresponding to the constant
transverse mode $q_{x}=0$. 

Noteworthy, even if the formal expressions given by Eqs.~(\ref{eq:w1-1},\ref{eq:w2-1})
are correct, they shall not be evaluated numerically using the inverse
$\mathbf{Q}^{-1}$ which can be numerically unstable. In the case
of a purely diffractive case, in general, a power expansion would
yield a dangerous method to evaluate a matrix exponential, see \cite{M-SIAMR-78}
for review on the matrix exponential and the discussion on the possibility
of a catastrophic cancellation. Notwithstanding, we found the Taylor
expansion to work well for our particular problem ---when there is
no index guiding.-- and to converge after only a few terms, some feature
which we attribute mainly to the small values of the $\left(\Delta_{0}h\right)$
pre-factor in the definition of $\mathbf{w_{0}}$. Here, the $\text{\ensuremath{\mathbf{w_{i,j}}}}$
can be conveniently evaluated from their series expansion that read
\begin{eqnarray}
\mathbf{w_{1}} & = & \sum_{n=0}^{\infty}\frac{h^{n+1}}{\left(n+1\right)!}\mathbf{Q}^{n},\\
\mathbf{w_{2}} & =\mathbf{w_{1,j}}- & \sum_{n=0}^{\infty}\frac{h^{n+1}}{\left(n+2\right)n!}\mathbf{Q}^{n}.
\end{eqnarray}
Very narrow gain stripes and/or large mesh discretization for which
$h\Delta_{0}\sim N_{x}^{-2}$ could however be problematic with this
simple method. 

In the more complex cases with constant losses and transverse index
guiding profiles, analytical estimates of $\mathbf{w_{1:2,j}^{\pm}}$
can still be obtained upon diagonalization, still without using any
potentially dangerous matrix inversion. This stem from the fact that
in this case $\mathbf{w_{0}^{\pm}}\left(s\right)=\exp\left(s\mathbf{Q}\right)=\mathbf{K^{-1}}\exp\left(s\mathbf{D}\right)\mathbf{K}$
with $\mathbf{K}$ an orthogonal matrix such that $\mathbf{K}^{-1}=\mathbf{^{t}K}$
and and $\mathbf{D}$ diagonal. Although $\mathbf{Q}$ is complex,
this diagonalization via an orthogonal matrix is always possible because
$\mathbf{Q}\sim i\left(\mathbf{M}+\mathbf{P}\right)$ with both $\mathbf{M}$
and $\mathbf{P}$ real symmetric. The fact that $\mathbf{P}$ is real
symmetric is not entirely trivial. We assumed the index profile to
be real and symmetric, hence the Fourier transforms of the two functions
involved in the construction of $\mathbf{P}$, either $\psi\left(x\right)$
for the FT or $\left(1+ix\right)\psi\left(x\right)$ for the RCT,
are real symmetric, in the latter case because the Fourier transform
of $x\psi\left(x\right)$, and odd real function, is purely imaginary.

\bibliographystyle{unsrt}
\bibliography{../BIBLIO/full}

\end{document}